\documentclass[structabstract]{aa} 
\usepackage{graphicx}
\usepackage{amssymb}
\usepackage{subfigure}
\usepackage{txfonts}
\usepackage{natbib}
\begin{document}

   \title{A search for star-planet interactions in the $\upsilon$~Andromedae system at X-ray and optical wavelengths}

   \author{K.~Poppenhaeger\inst{1}
      \and
       L.~F.~Lenz\inst{2}
      \and
       A.~Reiners\inst{2}    
      \and
       J.~H.~M.~M.~Schmitt\inst{1}
      \and
       E.~Shkolnik\inst{3}}
   \institute{Hamburger Sternwarte, Hamburg University,
             Gojenbergsweg 112, 21029 Hamburg, Germany\\
              \email{katja.poppenhaeger@hs.uni-hamburg.de}
	          \and
		  Universit\"at G\"ottingen, Institut f\"ur Astrophysik, Friedrich-Hund-Platz 1, 37077 G\"ottingen, Germany
		  \and
		  Carnegie Institution of Washington, Department of Terrestrial Magnetism, 5241 Broad Branch Road N.W., Washington, DC 20015
             }

   \date{Received 27 October 2010; accepted 08 January 2011}

  \abstract
{Close-in, giant planets are expected to influence their host stars via tidal or magnetic interaction. But are these effects in X-rays strong enough in suitable targets known so far to be observed with today's instrumentation?}{The $\upsilon$~And system, an F8V star with a Hot Jupiter, was observed to undergo cyclic changes in chromospheric activity indicators with its innermost planet's period. We aim to investigate the stellar chromospheric and coronal activity over several months.}{We therefore monitored the star in X-rays as well as at optical wavelengths to test coronal and chromospheric activity indicators for planet-induced variability, making use of the \em Chandra \rm X-ray Observatory as well as the echelle spectrographs \em FOCES \rm and \em HRS \rm at Calar Alto (Spain) and the Hobby-Eberly Telescope (Texas, US).}{The stellar activity level is low, as seen both in X-rays as in \ion{Ca}{ii} line fluxes; the chromospheric data show variability with the stellar rotation period. We do not find activity variations in X-rays or in the optical that can be traced back to the planet.}{Gaining observational evidence of star-planet interactions in X-rays remains challenging.}

   \keywords{Planet-star interactions -- Stars: activity -- Stars: coronae -- Stars: chromospheres -- X-rays: stars -- Stars: individual: upsilon Andromedae -- X-rays: individuals: upsilon Andromedae }

\titlerunning{star-planet interactions in the $\upsilon$~Andromedae system}

\maketitle

% \abstract{}{}{}{}{} % 5 {} token are mandatory

%_______________________________________________________________

\section{Introduction}

Interactions between stars and their giant planets have received considerable attention during the last years. Since the discovery of 51~Peg~b, a Jupiter-like planet in a four-day orbit around a solar-like star, the influence of stellar irradiation on planetary atmospheres has been investigated by various authors \citep[cf.][]{Lammer2003, Lecavelier2007, Erkaev2007}, as have possible effects that giant planets may have on their host star's activity. These star-planet interactions (SPI) are thought to happen via one of two major  scenarios \citep{CuntzSaar2000}, either tidal or magnetic interaction. In the tidal interaction process, the planet induces tidal bulges on the surface of the star, which can cause enhanced stellar activity through increased turbulence in the photosphere. Magnetic interactions can increase stellar activity via magnetic reconnection of planetary and stellar magnetic field lines \citep{Lanza2008}, or via Jupiter-Io-like interaction, i.e. flux tubes that connect star and planet and heat the stellar atmosphere at their footpoints \citep{Goldreich1969, Schmitt2009Io}.

There have been dedicated searches for observational signatures of SPI in stellar chromospheres. \cite{Shkolnik2005} investigated 13 stars and found for two stars variations in the \ion{Ca}{ii} K line core fluxes, a common chromospheric activity indicator, which were in phase with the planetary orbit. That flux excesses appeared once per orbit and not twice suggests that magnetic and not tidal interaction might have been at work. A follow-up study \citep{Shkolnik2008} found variability with the planetary orbit only for one of the targets during 2005; for other observation times and other targets, only a variability with the stellar rotation period was found. This was interpreted as on/off behavior of SPI; theoretical considerations also show that possible SPI signatures can be variable in time given the changing magnetic configurations in stellar atmospheres \citep{CranmerSaar2007}.

Other studies extended the search for SPI signatures into the X-ray regime to test for coronal activity changes in addition to the chromospheric hints found before. A statistical study by \cite{kashyapdrakesaar2008} claimed that stars with close-in planets are over-active by a factor of four in X-rays, but these authors had to use a large number of upper limits. \cite{Poppenhaeger2010} investigated a complete sample of all planet-bearing stars within 30~pc distance and found no correlation between stellar activity and planetary distance or mass that could not be traced back to selection effects. \cite{Scharf2010} investigated a smaller sample of planet-hosting stars at distances of up to $60$~pc and found a correlation of stellar X-ray luminosity and planetary mass for very close planets; however, selection effects could not be ruled out for the low-mass planets ($M_p < 0.1 M_J$) in the sample.

We here investigate the $\upsilon$~And (HD~9826) system for SPI signatures in chromospheric as well as coronal data. The system is one of the star-planet systems with available observations of time-variable (in terms of on/off behavior) SPI signatures as published in \cite{Shkolnik2005, Shkolnik2008}. $\upsilon$~And is an F8V star, orbited by a massive planet ($0.69 M_J$) in a $4.6$~d ($a=0.059$~AU) orbit, as well as by two other planets at much larger distances. This makes $\upsilon$~And a very suitable candidate to search for SPI signatures.

To look for signatures of SPI, we observed the $\upsilon$~And system for the first time nearly simultaneously at optical and X-ray wavelengths, thus testing for chromospheric as well as coronal variability. Because stellar activity in general and expected SPI signatures in particular do not have to be constant in time, it is important to observe the stellar chromosphere and corona at close time epochs.

\section{Observations and data analysis}\label{analysis}

Our optical data were obtained with the FOCES echelle spectrograph \citep{Pfeiffer1998} at Calar Alto, Spain, and with HRS \citep{Tull1998}, the echelle spectrograph mounted at HET in Texas, US. The X-ray data were collected with the Chandra X-ray Observatory. A complete list of our observations is given in Table~\ref{observations}.

%%%%%%%%%%%%%%%%%%%%%%%%%%%%%%%%%%%
   \begin{table}
      \caption[]{Optical and X-ray observations of $\upsilon$~And, sorted by observation date. Integrated \ion{Ca}{ii}~K line residuals are given for optical data, but the absolute scales of FOCES and HRS data are not comparable (see text); extrapolated SPI state according to \cite{Shkolnik2005} given for X-ray data (see text).}
        \label{observations}
    \begin{tabular}{l l r l}
    \hline\hline
    Instrument 		& MJD	& \ion{Ca}{ii} K residual 	& exp. SPI state \\ \hline
    FOCES (15$\mu$) & 55014.12& $   -0.05\pm    0.14$ & - \\
    FOCES (15$\mu$) & 55015.12& $   -0.03\pm    0.19$ & - \\
    FOCES (15$\mu$) & 55016.12& $   -0.05\pm    0.11$ & - \\
    FOCES (15$\mu$) & 55017.12& $   -0.01\pm    0.11$ & - \\
    FOCES (15$\mu$) & 55018.12& $    0.10\pm    0.14$ & - \\
    FOCES (15$\mu$) & 55020.13& $    0.36\pm    0.14$ & - \\
    FOCES (15$\mu$) & 55021.13& $    0.07\pm    0.11$ & - \\
    FOCES (15$\mu$) & 55022.13& $   -0.03\pm    0.13$ & - \\
    FOCES (15$\mu$) & 55023.13& $   -0.39\pm    0.20$ & - \\
    FOCES (15$\mu$) & 55024.14& $   -0.24\pm    0.14$ & - \\
    FOCES (15$\mu$) & 55025.13& $   -0.13\pm    0.14$ & - \\
    FOCES (15$\mu$) & 55027.13& $    0.36\pm    0.09$ & - \\
    FOCES (15$\mu$) & 55028.13& $    0.37\pm    0.09$ & - \\
    FOCES (15$\mu$) & 55029.12& $    0.27\pm    0.14$ & - \\
          HET / HRS & 55049.37& $    0.50\pm    0.54$ & - \\
          HET / HRS & 55083.27& $   -0.05\pm    0.49$ & - \\
          HET / HRS & 55090.25& $   -0.76\pm    0.53$ & - \\
    FOCES (24$\mu$) & 55096.19& $    0.03\pm    0.17$ & - \\
    FOCES (24$\mu$) & 55099.20& $   -0.08\pm    0.13$ & - \\
    FOCES (24$\mu$) & 55100.20& $   -0.05\pm    0.17$ & - \\
    FOCES (24$\mu$) & 55105.21& $   -0.02\pm    0.14$ & - \\
    FOCES (24$\mu$) & 55107.20& $   -0.01\pm    0.18$ & - \\
    FOCES (24$\mu$) & 55108.17& $   -0.01\pm    0.16$ & - \\
    FOCES (24$\mu$) & 55109.17& $   -0.08\pm    0.16$ & - \\
    FOCES (24$\mu$) & 55110.19& $   -0.17\pm    0.16$ & - \\
          HET / HRS & 55110.44& $   -1.10\pm    0.53$ & - \\
    Chandra ACIS-S	&55124.72	& -	& maximum \\
    Chandra ACIS-S	&55126.56	& -	& minimum \\
    Chandra ACIS-S	&55131.29	& -	& minimum \\
    Chandra ACIS-S	&55133.50	& -	& maximum \\ 
          HET / HRS & 55135.12& $   -0.22\pm    0.47$ & - \\
          HET / HRS & 55139.34& $   -0.43\pm    0.49$ & - \\
          HET / HRS & 55141.33& $    0.72\pm    0.47$ & - \\
          HET / HRS & 55142.35& $    1.00\pm    0.49$ & - \\
          HET / HRS & 55146.33& $    0.35\pm    0.48$ & - \\
    FOCES (24$\mu$) & 55169.00& $    0.13\pm    0.20$ & - 
    \end{tabular}
   \end{table}
%%%%%%%%%%%%%%%%%%%%%%%%%%%%%%%%%%%

\subsection{Optical data from FOCES}

We obtained 23 spectra of $\upsilon$~And with the FOCES echelle spectrograph installed at the 2.2m telescope at Calar Alto from July to December 2009, when one spectrum per night was taken. The 14 spectra obtained during July were recorded with the LOR\#11i detector with 15$\mu$m pixel size, the other 9 spectra were recorded with the Site\#1d detector with 24$\mu$m pixel size, all with an exposure time of $900$s. The covered wavelength range was 3900--9500~$\AA$.

The data were reduced with Calar Alto's standard echelle extraction routine d2.pro written in IDL (available from the Calar Alto data server ftp.caha.es). Mean dark frames were subtracted from the nightly spectra, which were then normalized using mean flat fields taken prior to the observations. Wavelength calibration was done with Thorium-Argon frames taken during the same night. The spectral resolution as inferred from the FWHM of ThAr lines is $R\sim 30000$. Since not all of our spectra have ThAr frames taken directly before or after the observation, we cross-correlated the spectra around the \ion{Ca}{ii} H and K lines and shifted the wavelength axis correspondingly. This deprives us of the possibility to study RV shifts, but does not interfere with our analysis of flux excesses in the \ion{Ca}{ii} H and K line cores.

\begin{figure}
\includegraphics[width=0.5\textwidth]{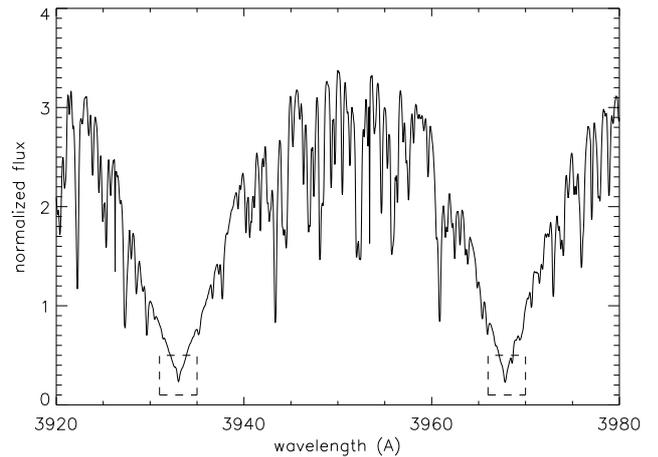}
\caption{Typical spectrum of $\upsilon$~And recorded with FOCES. The reversals in the \ion{Ca}{ii} H and K line cores (dashed boxes) are weak.}
\label{opticalspectrum}
\end{figure}

A typical spectrum of the \ion{Ca}{ii} H and K lines of $\upsilon$~And is shown in Fig.~\ref{opticalspectrum}. The reversals in the two line cores are weak, but clearly visible. Because of the very small emission peaks and since activity affects the K line profile more strongly than the H line profile (see for example \cite{HallLockwoodSkiff2007}), we concentrate on the K line here. The signal-to-noise ratio (S/N) of the final spectra is at $\approx150$~pixel$^{-1}$ ($\approx 860\,\AA^{-1}$) in the quasi-continuum between the H and K line and at $\approx40$~pixel$^{-1}$ ($\approx 230\,\AA^{-1}$) in the line cores, summing up to $S/N\approx 150$ in the complete line cores, which stretch over ca.~33 pixels each.

%%%%%%%%%%%%%%%%%%%%%%%%%%%%%%%%%%%
%%%%%%%%%%%%%%%%%%%%%%%%%%%%%%%%%%%

The spectra were normalized by setting the flux of one template spectrum at 3929.5~$\AA$ to unity, giving a comparable normalization to the one used by \cite{Shkolnik2005}. The remaining spectra were normalized by minimizing the scatter of the flux in two areas to the left and right of the \ion{Ca }{ii}~K line reversal, specifically in the 3930-3932$\AA$ and 3934-3936$\AA$ range. Later on, a mean spectrum was subtracted from the individual spectra to determine the flux variations in the line cores (see section~\ref{chrom}). We tested that shifting the normalization areas by 1-2$\AA$ outward from the line cores does not significantly change our results. To further test if our normalization method might introduce false variability signals, we also applied our procedure to the \ion{Al}{i} line near 3944$\AA$, a photospheric line that should show no activity-related variations. In that part of the spectra, the multitude of lines present makes the normalization less exact, causing a higher overall scatter between the normalized spectra, specifically a noise level of $1 \sigma \approx 0.03$ in the chosen normalization. However, the magnitude of the residuals of the individual spectra compared with the mean spectrum does not change near the \ion{Al}{I} line, although it has a comparable depth to the \ion{Ca}{ii} lines, indicating that the method itself does not produce false variability signals bet\-ween the two normalization areas.

For the subsequent analysis, the residuals at the \ion{Ca}{ii}~K line core were integrated over a central range of $1\AA$. The corresponding error bars were calculated by estimating the statistical error in each bin to be the standard deviation of the residuals in the normalization ranges, and then calculating the 2$\sigma$ error in the integrated area under the \ion{Ca}{ii} K line center by Monte-Carlo simulations.

\subsection{Optical data from HRS}\label{hetreduction}

Nine spectra were taken with the HRS instrument at the HET with a resolution of $R\sim 60000$. The data reduction was similar to the FOCES data reduction except for the use of flatfields. The HRS flatfields are taken with a separate, wider calibration fiber to allow 2D-flatfielding. However, the orders of the calibration fiber overlap in the \ion{Ca}{ii}~K region of the flatfields and thus could not be used. We tried to use standard star data as a surrogate for a flatfield, but this added a lot of noise because of their relatively low data quality. To preserve the signal-to-noise ratio of the spectra ($S/N\approx90$~pixel$^{-1}\approx 520\,\AA^{-1}$ in the \ion{Ca}{ii}~K core) we reduced the data without flatfielding.

A consequence of the lacking flatfields is that a direct comparison between the data from FOCES and HRS is not possible. We experimented with artificial, polynomial fits to the continua to obtain comparable line shapes in the \ion{Ca}{ii}~K line, but did not achieve an acceptable solution. As a result the two data sets had to be analyzed individually.

The HRS spectra were normalized by first setting their flux at 3929.5~\AA~ to unity. Then a reference spectrum was chosen and a polynomial was fitted to the quotient of each spectrum and the reference spectrum that represents the missing flatfield. The \ion{Ca}{ii}~K line was omitted from the fitting to avoid removing any variability in the line core. Each spectrum was divided by its fit to receive the normalized spectra.

\subsection{X-ray data}

We observed the $\upsilon$~And system with the {\em Chandra} X-ray telescope in four pointings of $15$~ks ($\approx 4$~h) each. These observations were scheduled in a way that two of them cover the projected maximum and two the projected minimum SPI-activity times, as inferred from optical observations done by \cite{Shkolnik2005} (see Table~\ref{observations}). Because SPI can depend on numerous factors such as the stellar magnetic field configuration, SPI signatures are expected to be time-variable \citep{CranmerSaar2007, Shkolnik2008}. It is unlikely, consequently, that we will find SPI in exactly the same configuration as the system exhibited in the chromospheric data from 2005; however, our observational schedule of two observations each that are spaced apart one half of an orbital period is suited to uncover signatures of planet-induced coronal hot spots.

The data were reduced using standard procedures of the CIAO~v.4.2 software package. $\upsilon$~And emits soft X-ray radiation with practically all photons having energies below 2~keV, its mean X-ray count rate is $\approx 0.025$~cts/s in the 0.4-2~keV energy band. We produced light curves with $1$~ks binning to obtain acceptable error bars as well as sufficient time resolution to identify possible flares. For the spectra we used energy bins with at least 15 counts per bin for decent statistics. The spectral fitting was performed with Xspec~v12.5.

Because $\upsilon$~And is an optically rather bright star ($m_V = 4.09$), the data were collected with a reduced frame time of $0.74$s. Additionally, we only use data at energies above $400$~eV to exclude possible optical contaminations at the low-energy sensitivity end of the detector.

\section{Results}

\subsection{Chromospheric activity}\label{chrom}

%%%%%%%%%%%%%%%%%%%%%%%%%%%%%%%%%%%
\begin{figure}
\includegraphics[width=0.5\textwidth]{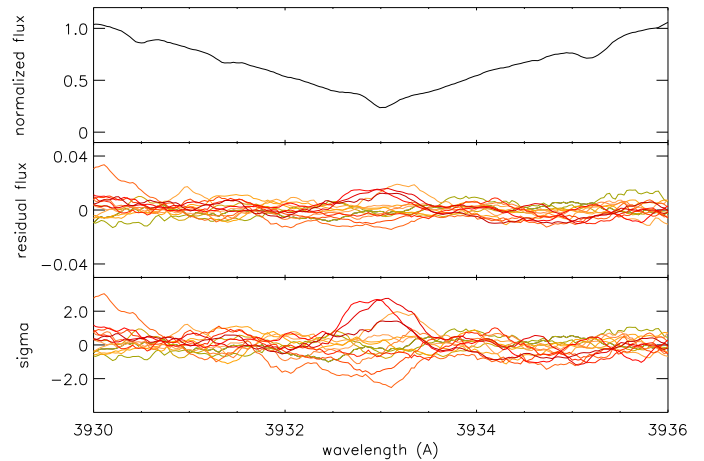}
\caption{Variability in \ion{Ca}{ii}~K line cores of the FOCES $15\mu$ data. {\em upper panel:} normalized mean spectrum; {\em middle panel:} residual flux with the same normalization; {\em lower panel:} flux variation in standard deviations.}
\label{residuals}
\end{figure}
%%%%%%%%%%%%%%%%%%%%%%%%%%%%%%%%%%%

%%%%%%%%%%%%%%%%%%%%%%%%%%%%%%%%%%%
\begin{figure}
\includegraphics[width=0.5\textwidth]{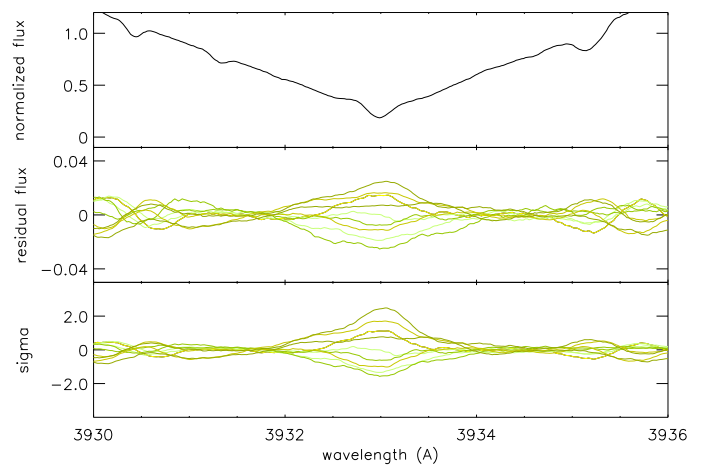}
\caption{Same as Fig.~\ref{residuals}, but for HRS data.}
\label{residuals_het}
\end{figure}
%%%%%%%%%%%%%%%%%%%%%%%%%%%%%%%%%%%

Following \cite{Shkolnik2005}, we computed separate median spectra for the FOCES and the HRS data and computed the residual fluxes of each spectrum compared with the respective median spectrum by subtracting the median spectra binwise from the individual spectra. We also computed the variation measured in standard deviations by dividing the residuals by the poissonian flux error of each bin of the individual spectra, neglecting the error in the median spectra. The results for the 15$\mu$ data, smoothed by 15 bins ($\approx 0.45\AA$), are shown in Fig.~\ref{residuals}. For the 15$\mu$ data, which were recorded in July~2009 under very favorable weather conditions, we find variations in the \ion{Ca}{ii}~K line core at $\approx 2\sigma$ levels; the HRS data show similar variation (Fig.~\ref{residuals_het}), while for the 24$\mu$ data, the overall noise level is so high that no additional variability in the line core can be identified by naked-eye inspection. 

The HRS residuals show variations over a region broader than the \ion{Ca}{ii}~K
line core. To the sides of the cores, the residuals also exhibit some
broad, wave-like structure. This pattern is probably caused by our
normalization; we had to use a polynomial fit to substitute the missing flatfields that apparently does not entirely capture the region around the
K~line. Nevertheless, the residuals at the K line core exceed the
variations outside the core, and therefore we believe that at least some of the
core variation is real. Varying the footpoints chosen for the polynomial
fit does not significantly change the residuals. Thus, the signal of the
K residuals is stable against normalization, but the total uncertainty
in the \ion{Ca}{ii}~K line core residuals may contain quite large errors. To estimate the magnitude of these errors, we note that the variation outside the K~line core is at a flux level of $\pm 0.01$; integrated over the 33~pixels line core, this yields an additional uncertainty of 0.33, which we add to our nominal errors of the integrated residuals.

The magnitude of the K~line fluctuations, measured in relative flux deviations, is at a $\pm 0.02$ level in our chosen normalization, with an overall noise level of $\pm 0.01$. \cite{Shkolnik2005} found for $\upsilon$~And a variation at a $\pm 0.01$ level and an overall noise level of $\pm 0.002$. Our noise level is much higher with $\pm 0.01$, and so we interpret our $\pm 0.02$ variation in the core to be consistent with the previously found level of $\pm 0.01$; however, given the time-dependence of activity features in stellar chromospheres, one could also expect a differing level of variability.

The \ion{Ca}{ii}~K line core fits into an interval of $\approx 1 \AA$ width, see for example \cite{HallLockwoodSkiff2007} and our data in Fig.~\ref{residuals}. We therefore integrate the K line residuals from 3932.5-3933.5$\AA$ to obtain a measure for the total variation per spectrum. The crucial question now is whether there is a periodicity in this signal, and if there is, which period can be associated with it.

\begin{figure}[t!]
\includegraphics[width=0.5\textwidth]{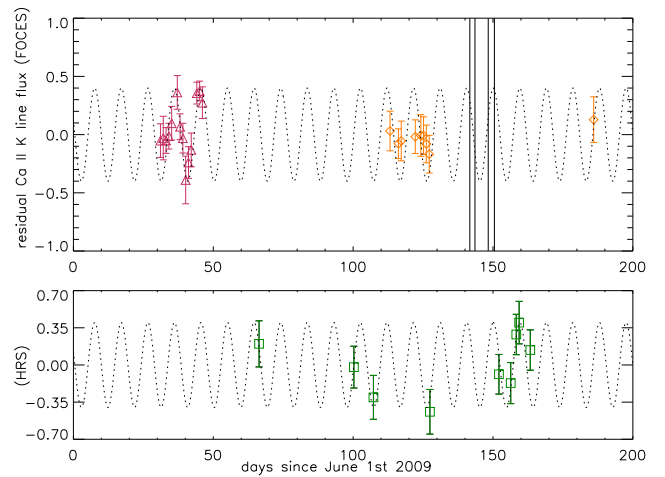}
\caption{Time series of \ion{Ca}{ii} residuals with the highest probability period ($9.5$~d) indicated by the dotted line; red triangles and orange diamonds are $15\mu$ and $24\mu$ data from FOCES, green boxes are scaled-down HRS data with the same periodicity indicated. {\em Chandra} observation dates are indicated by vertical solid lines.}
\label{period_model}
\end{figure}

\begin{figure}[t!]
\includegraphics[width=0.5\textwidth]{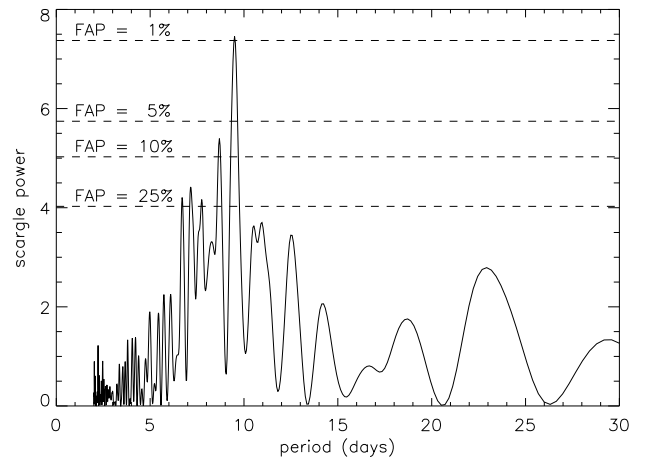}
\caption{Lomb-Scargle periodogram of \ion{Ca}{ii}~K line residuals (weighted by their respective errors) with nominal false-alarm probabilities given by the horizontal lines (see text).}
\label{period}
\end{figure}

The timeseries of our integrated residuals is shown in Fig.~\ref{period_model}. The FOCES 15$\mu$ data show already by naked-eye inspection a variability that tracks slightly more than one cycle of a presumably sinusoidal variation with a periodicity of about eight to nine days. As discussed in section~\ref{hetreduction}, the spectra from FOCES and HET cannot be absolutely calibrated with respect to each other, and therefore the calculated residuals are not comparable on an absolute scale either. Thus we proceeded as follows: First, we calculated a Lomb-Scargle periodogram \citep{Lomb1976, Scargle1982} of the collected optical data from FOCES, weighting the data points with their respective errors \citep[see][]{GillilandBaliunas1987}. This yields three significant peaks with false-alarm probabilities (FAP) below $5\%$, corresponding to periods of $P=9.3$~d, $8.7$~d, and $8.2$~d, sorted by descending significance. Then, we calculated a weighted Lomb-Scargle periodogram for the HRS data alone. This yields no significant periods with FAP below $20\%$, which could be expected because there are only nine data points from HRS, distributed over more than three months. Finally, we tried to combine the HRS and FOCES residuals {\it ad hoc} by scaling down the HRS residuals by a factor of $0.4$ so that the highest and lowest values approximately match in both datasets. This is a crude approximation at best; in the Lomb-Scargle periodogram, a possibly wrong scaling will lead to incorrect FAPs, but prominent peaks should still be reproduced reliably. We note here that this scaling also roughly fits with the additional variation level outside the line core that is present in the HRS data mentioned above. The result of our periodicity test is shown in Fig.~\ref{period}. The periodogram of the combined data exhibits the strongest peak at $P=9.5$~d with a nominal FAP of $0.9\%$ (keeping in mind that the FAPs may be unreliable for the combined data set). The important point here is that we see a single, isolated main peak at a period that is also found in the FOCES data alone. In contrast to this, near the planetary orbital period of $4.6$~d, or half its value, no significant peak appears in any of the periodograms. The periodicities are also clearly reflected in Fig.~\ref{phase_star} and \ref{phase_planet}, where we show the variability of the residuals over the stellar rotation period and the planetary orbital period.

%%%%%%%%%%%%%%%%%%%%%%%%%%%%%%%%%%%
%%%%%%%%%%%%%%%%%%%%%%%%%%%%%%%%%%%
%%%%%%%%%%%%%%%%%%%%%%%%%%%%%%%%%%%
%%%%%%%%%%%%%%%%%%%%%%%%%%%%%%%%%%%

For testing purposes, we subtracted the $9.5$~d periodicity from our data to see if the remaining residuals exhibit other periodicities (such as with the orbital period). However, a periodogram of these residuals does not yield other significant periods. Given the expected variability of SPI signatures with time \citep{Shkolnik2008}, we also tried searching for periodicities in subgroups of our data, but the number of data points is not high enough to allow for significant period detection then.

If one chooses to interpret these findings as signatures of periodic activity variations, they are probably associated with the stellar rotation period. The star's rotational velocity $v\sin i$ was measured to be $9.5\pm 0.4$~km/s by \cite{Gonzales2010}, its radius is computed by \cite{HenryBaliunas2000} from stellar parameters as $1.6$R$_{\sun}$, yielding a rotation period of $\approx8.5$~d, with rather large, but difficult to quantify errors, since the stellar radius was not determined observationally. \cite{WrightMarcy2004} give a rotational period of $12$~d from spectroscopic monitoring; \cite{HenryBaliunas2000} find only weak signatures of rotational modulation with periods of $11$~d and $19$~d respectively in two different data sets. They also state that the difference of the spectroscopically derived periods to the estimate derived from $v\sin i$ measurements might be caused by differential rotation. Still, this stellar rotation period fits the possible periodic signal in our data better than the orbital period of the Hot Jupiter of 4.6~d. Additionally, we have a subset of our data consisting of nightly measurements in July 2010, which closely tracks one complete sinusoidal variation of $\approx 9$~d period, making it rather unlikely that we see an alias of the planetary period here, but not the period of $4.6$~d itself. This suggests that we see typical low-level stellar activity variations with the stellar rotation period that are not induced by SPI.

\subsection{Coronal activity}

\begin{figure}
\includegraphics[width=0.5\textwidth]{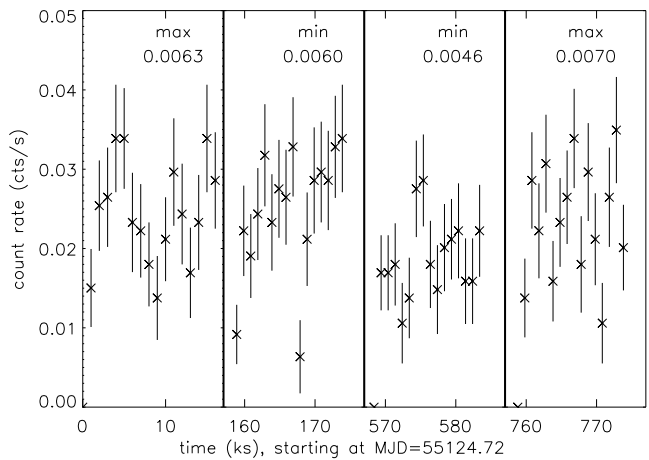}
\caption{Background-subtracted X-ray lightcurves of $\upsilon$~And, taken with {\em Chandra} ACIS-S in the 0.4-2.0 keV energy band, with expected SPI states according to \cite{Shkolnik2005} and MAD values indicated.}
\label{lightcurves}
\end{figure}

We extracted X-ray lightcurves of $\upsilon$~And with 1~ks binning in the 0.4-2.0~keV energy band. The lightcurves (see Fig.~\ref{lightcurves}) show variability at 50\% level, but no large flares. The mean count rate is constant in observations 1, 2, and 4; the third observation's mean count rate is somewhat lower by 25\%. Applying the concept of mean average deviation (MAD) that was used in \cite{Shkolnik2008}, we find that the MADs of all but the third observation are similar ($0.0063$, $0.0060$ and $0.0070$), whereas the third observation has a MAD that is also lower by $\approx 25\%$ compared with the rest ($0.0046$).

\begin{figure}
\includegraphics[width=0.35\textwidth,angle=270]{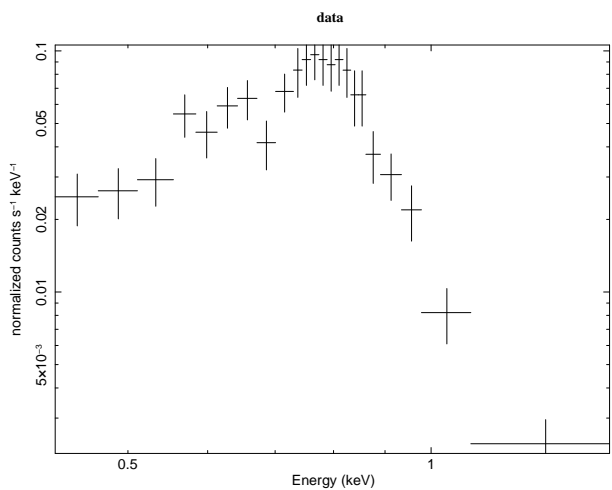}
\caption{{\em Chandra} ACIS-S spectrum of $\upsilon$~And extracted from a single 15~ks exposure. Strong emission in the iron line complexes around 800~eV is visible. }
\label{Xrayspec}
\end{figure}

A typical X-ray spectrum of $\upsilon$~And is shown in Fig.~\ref{Xrayspec}; these spectra have a total amount of $\approx 450$ source counts, binned by 15 counts as a minimum to allow $\chi^2$ statistics. The spectra of all four pointings cannot be satisfactorily fitted with thermal plasma models with solar abundances and one or two temperature components, while a one-temperature model with variable elemental abundances yields acceptable fits. The results of the spectral fitting performed in Xspec~v.12.5 are given in Table~\ref{Xspec}; the abundances are given with regard to \cite{grevessesauval1998}. The modeled elemental abundances are interdependent with the derived emission measure, and different combinations of both parameters lead to very similar results. For example, the best fits of the first and last observation give lower abundances and higher emission measures than the fits of the other two pointings, but fixing the abundances to the values of the other observations yields comparable emission measures and almost the same fit quality.

%%%%%%%%%%%%%%%%%%%%%%%%%%%%%%%%%%%
\begin{figure}
\includegraphics[width=0.5\textwidth]{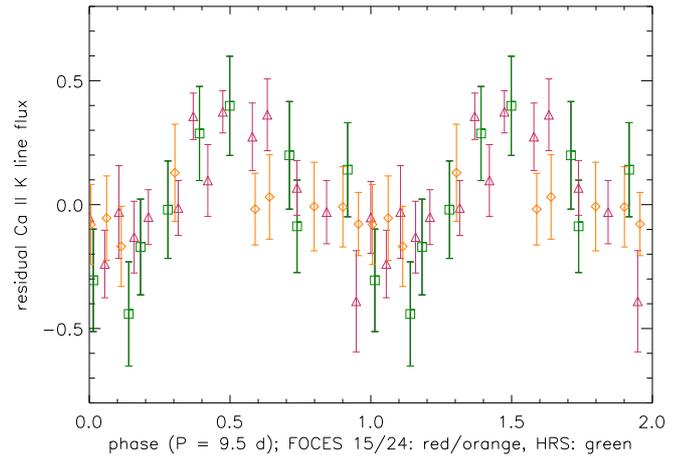}
\caption{\ion{Ca}{ii} K line residuals phase-folded with a period of $9.5$~d, presumably the stellar rotation period. Red triangles and orange diamonds are FOCES $15\mu$/$24\mu$ data, green squares are HRS data.}
\label{phase_star}
\end{figure}
%%%%%%%%%%%%%%%%%%%%%%%%%%%%%%%%%%%
%%%%%%%%%%%%%%%%%%%%%%%%%%%%%%%%%%%
\begin{figure}
\includegraphics[width=0.5\textwidth]{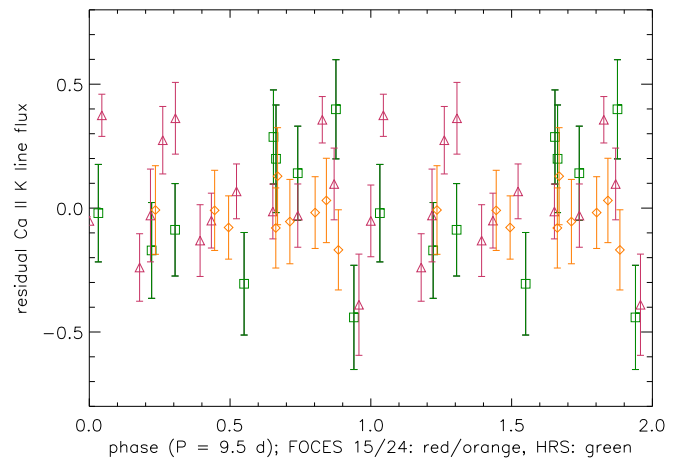}
\caption{Same as Fig.~\ref{phase_star}, but folded with the planetary orbital period of $4.6$~d.}
\label{phase_planet}
\end{figure}
%%%%%%%%%%%%%%%%%%%%%%%%%%%%%%%%%%%

%%%%%%%%%%%%%%%%%%%%%%%%%%%%%%%%%%%
   \begin{table*}
   \begin{center}
      \caption[]{Spectral modeling results with $1\sigma$ errors; emission measure given in units of $10^{50}$~cm$^{-3}$.}
        \label{Xspec}
    \begin{tabular}{l l l l l l}
    \hline\hline
    Parameter	 		&obs. 1		& obs. 2	& obs. 3	& obs.4 	& all \\ \hline
    $T_1$ (MK)			&$2.7\pm0.1$	&$3.1\pm0.1$	&$3.0\pm0.2$	&$2.8\pm0.1$	&$2.9\pm0.1$\\
    $EM_1$	 		&$8.1\pm1.6$	&$4.4\pm0.7$	&$4.1\pm0.9$	&$6.2\pm1.0$	&$5.2\pm0.5$\\
    $O$	 			&$0.15\pm0.06$	&$0.34\pm0.10$	&$0.24\pm0.09$	&$0.16\pm0.05$	&$0.23\pm0.04$\\
    $Ne$	 		&$0.22\pm0.08$	&$0.26\pm0.10$	&$0.25\pm0.10$	&$0.18\pm0.09$	&$0.23\pm0.05$\\
    $\chi^2_{red}$ (d.o.f.)	&0.81 (21)	&0.79 (20)	&0.78  (14)	&1.47 (19)	&1.19 (86) \\ \vspace{0.2cm}
    expected SPI state		&maximum	&minimum	&minimum	&maximum	& - \\
    $\log L_X$ (0.25-2.0 keV)	& 27.80		&27.62		&27.56		&27.69		& 27.65\\  \hline
    \end{tabular}
   \end{center}
   \end{table*}
%%%%%%%%%%%%%%%%%%%%%%%%%%%%%%%%%%%
% emission measure: upsand distance = 13.5pc=4.166e19cm -> 4.58e-55 * EM = Xspec_norm

Within errors, the spectral properties of all four observations are similar. The plasma temperature is fairly low with $\approx 3$~MK. The elemental abundances show a FIP effect, because elements with high first ionization potentials such as oxygen and neon are underabundant compared with iron with a low FIP. This is typical for stars with low to moderate coronal activity, which is determined by the activity indicator $\log L_X/L_{bol} < -4$. $\upsilon$~And's mean X-ray luminosity in these observations is $27.6$~erg\,s$^{-1}$, its bolometric luminosity is calculated according to \cite{flower1996} from $m_V=4.09$ and $B-V=0.54$ to be $3.3\,L_{bol\sun}$ and therefore its activity indicator is $L_X/L_{bol}=-6.5$, marking $\upsilon$~And as a fairly inactive star.

If the chromospheric data really track the stellar rotation, one might expect to see variability with that period in X-rays as well. This is different from the expected minima and maxima mentioned in Table~\ref{observations}, since we are now dealing with the chromospheric variability from our nearly-simultaneous optical data with a period of $9.5$~d. We indicated the times of the {\em Chandra} pointings as solid vertical lines in Fig.~\ref{period_model}; they correspond to rotational phases of $0.64$, $0.83$, $0.33$, and $0.57$ as given in Fig.~\ref{phase_star}. This means that the first, third, and last observation took place at times where the chromospheric activity was (by comparison) high, while the second observation was conducted at moderate chromospheric activity. The mean count rate and mean X-ray luminosity is lower in observation three, but observation two yields values comparable with the first and last observation. We conclude that the coronal activity seems to be dominated by short-term statistical variations and not by the periodicity seen in the chromospheric data.

%%%%%%%%%%%%%%%%%%%%%%%%%%%%%%%%%%%
%%%%%%%%%%%%%%%%%%%%%%%%%%%%%%%%%%%

%%%%%%%%%%%%%%%%%%%%%%%%%%%%%%%%%%%
%%%%%%%%%%%%%%%%%%%%%%%%%%%%%%%%%%%

\section{Discussion}

Searching for signatures of SPI in stellar coronae has proven to be a subtle task. Initial chromospheric measurements indicated that HD~179949's and $\upsilon$~And's chromospheric \ion{Ca}{ii} fluxes varied with the respective planetary period, but follow-up observations detected dominant variability with the stellar rotation period for several observational epochs. Previous attempts to observe possible SPI signatures in X-rays yielded detections of some activity features such as flares or elevated mean countrates, but attributing these effects unambiguously to SPI is difficult \citep{Pillitteri2010, SaarSPI2008}. 

The $\upsilon$~And system was one of the prime suspects for observing SPI at work in individual star-planet systems based upon chromospheric observations \citep{Shkolnik2005}. However, our observations of the system do not show any significant variations that could be attributed to planetary effects. The variations in the chromospheric \ion{Ca}{ii}~K line cores are small and are consistent with the stellar rotation period. The magnitude of the variations is $\approx 0.6\%$ of the flux in the pseudo-continuum between the \ion{Ca}{ii}~H and K line. From optical and X-ray monitoring of stars such as 61~Cyg \citep{HempelmannRobrade2006} we know that changes in the S-index (counts in \ion{Ca}{ii}~H and K lines normalized by counts in continuum stretches) of $\pm 15\%$ translate to changes in the $0.2-2.0$~keV X-ray band of $\pm 40\%$. Accordingly, the extremely small chromospheric variations of $\upsilon$~And should, if ruled by the same activity effects, cause coronal variations of less than two percent over one stellar rotation period. This is much lower than the typical intrinsic variation level of a late-type star. One would only expect strong SPI signatures in X-rays here if the SPI mechanism is fundamentally different from normal activity processes, for example, if SPI happened via Jupiter-Io-like interactions where the star and its close planet are connected by flux tubes (which cannot be the case for the $\upsilon$~And system since the stellar rotation period is longer than the planetary orbital period). If therefore SPI signatures are ruled by similar processes as general stellar activity, it is expected that for low-activity stars possible X-ray SPI effects can hide in the intrinsic stellar variability level. For the corona and the chromosphere of $\upsilon$~And we see that the star does not show any signs of planet-induced activity at the epoch of our observations; it is instead a low-activity star with some indication for rotational modulation in chromospheric emissions.

From this data and other searches for SPI signatures, it seems that stars with low to moderate activity do only exhibit very low levels of SPI effects. To unambiguously detect SPI signatures in the future, stars with extremely close-in planets ($<2$~d) will be the most promising candidates. According to recent models \citep{Lanza2009}, magnetic SPI can occur not only through reconnection between the stellar and planetary magnetic field lines, but also by the planetary field that disturbs stellar magnetic loops that have stored energy by normal stellar activity processes and triggers the release of energy. 
The analysis of observational data \citep{Shkolnik2008} has also shown that signatures of SPI may be detectable only at certain time epochs, presumably depending on the changing configuration of stellar and planetary magnetic fields. Stars with higher activity levels and larger coronal loops could be rewarding targets for SPI searches if observed with good phase coverage and higher $S/N$ to enable a differentiation between intrinsic stellar variability and SPI effects.

\section{Conclusions}
Our main results are summarized as follows:
\begin{enumerate}
\item $\upsilon$~And is a star of low chromospheric activity, with the coronal activity consistently being at a  low level as well, indicated by the coronal activity indicator $\log L_X/L_{bol} = -6.5$.
\item Our data show variations of the \ion{Ca}{ii}~K line core emission compared to the mean spectrum. These variations have a period of $\approx 9.5$~d, close to the stellar rotation period of $\approx 8.5$~d.
\item The X-ray data do not show significant changes between expected SPI maximum and minimum states, neither in the lightcurves nor in the spectra. The spectra show a FIP effect, with iron being overabundant by a factor of $\approx 4$ compared to neon, typical for stars with low to moderate X-ray activity.
\item In our observations, the $\upsilon$~And system does not show signatures of star-planet interactions. The periodicity observed in chromospheric activity indicators is very close to the calculated stellar rotation period and is therefore probably induced by non-SPI-related active regions on the star. Gaining observational evidence for star-planet interactions in X-rays remains a challenge.
\end{enumerate}

\begin{acknowledgements}
K.~P.~acknowledges financial support from DLR grant 50OR0703. A.~R.~acknowledges research funding from the DFG under RE~1664/4--1. This work is based on observations obtained with {\em Chandra} X-ray Observatory, {\em FOCES} at Calar Alto and {\em HRS} at the Hobby-Eberly Telescope, Texas. The Hobby-Eberly Telescope (HET) is a joint project of
the University of Texas at Austin, the Pennsylvania State
University, Stanford University, Ludwig-Maximilians-Universit\"at
M\"unchen, and Georg-August-Universit\"at G\"ottingen. The HET is
named in honor of its principal benefactors, William P. Hobby and
Robert E. Eberly.
\end{acknowledgements}

% bibtex in der Konsole ausführen
\bibliographystyle{aa}
\bibliography{/data/hspc74/Files/katjasbib.bib}

\end{document}